\documentclass[prl,floats,twocolumn,showpacs,floatfix,superscriptaddress]{revtex4}
\usepackage{graphics,graphicx,dcolumn,bm,fleqn,epic,eepic,float}
\usepackage{amssymb,amsmath,multirow,rotate,color,float}
\usepackage[latin1]{inputenc}

\usepackage[dvips]{epsfig}

\begin{document}
\title{Random-roughness hydrodynamic boundary conditions}

\author{Christian Kunert}
\affiliation{Institute for Computational Physics, University of Stuttgart, Pfaffenwaldring 27, 70569 Stuttgart, Germany}

\author{Jens Harting}
\affiliation{Department of Applied Physics, Eindhoven University of Technology, P.O. Box 513, 5600 MB Eindhoven, The Netherlands}
\affiliation{Institute for Computational Physics, University of Stuttgart, Pfaffenwaldring 27, 70569 Stuttgart, Germany}

\author{Olga I. Vinogradova}
\affiliation{A.N.~Frumkin Institute of Physical
Chemistry and Electrochemistry, Russian Academy of Sciences, 31
Leninsky Prospect, 119991 Moscow, Russia}
\affiliation{ITMC and DWI, RWTH Aachen, Pauwelsstr. 8,
52056 Aachen, Germany}


\date{\today}

\begin{abstract}
We report results of lattice Boltzmann simulations of a high-speed drainage of
liquid films squeezed between a smooth sphere and a randomly rough plane. A
significant decrease in the hydrodynamic resistance force as compared with that
predicted for two smooth surfaces is observed. However, this force reduction
does not represent slippage.
The computed force is exactly the same as that between equivalent smooth
surfaces obeying no-slip boundary conditions, but located at an intermediate
position between peaks and valleys of asperities. The shift in hydrodynamic
thickness is shown to depend on the height and density of roughness elements.
Our results do not support some previous experimental conclusions on very large
and shear-dependent boundary slip for similar systems.
\end{abstract}

\maketitle

{\bf Introduction.--}
It has been recently well recognized that the famous no-slip boundary
condition, for more than a hundred years applied to model experiments in
fluid mechanics, reflected mostly a macroscopic character and
insensitivity of old style experiments. Modern experiments concluded that
although the no-slip postulate is valid for molecularly smooth hydrophilic
surfaces down to
contact~\cite{vinogradova:03,charlaix.e:2005,vinogradova.oi:2009},
for many other systems it does not apply when the size of a system is reduced
to micro- and nano scales. The changes in hydrodynamic behavior are caused by
an impact of interfacial phenomena, first of all hydrophobicity and roughness,
on the flow.
The effect of hydrophobicity on the flow past smooth surfaces is reasonably
clear and suggests an amount of slippage described by the condition
$v_s = b \partial v / \partial z$
where $v_s$ is the slip velocity at the wall, $b$ the slip length, and the axis
$z$ is normal to the surface. The assumption is justified
theoretically~\cite{vinogradova:99,barrat:99,andrienko.d:2003,bib:jens-kunert-herrmann:2005}
and was confirmed by surface force apparatus
(SFA)~\cite{charlaix.e:2005}, atomic force microscope
(AFM)~\cite{vinogradova:03}, and fluorescence cross-correlation
(FCS)~\cite{vinogradova.oi:2009} experiments. Despite
some remaining controversies in the data and amount of slip
(cf.~\cite{lauga2007}), a concept of hydrophobic slippage is now widely
accepted.
If a liquid flows past a rough hydrophobic (i.e. superhydrophobic) surface,
roughness may favor the formation of trapped gas bubbles, resulting in a large
slip
length~\cite{bib:joseph_superhydrophobic:2006,ou2005,feuillebois.f:2009,sbragaglia-etal-06,kusumaatmaja-etal-08b,jari-jens-08}.
For rough hydrophilic surfaces the situation is much less clear, and
opposite experimental conclusions have been made: one is that roughness
generates extremely large slip~\cite{bonaccurso.e:2003}, and one is
that it decreases the degree of slippage~\cite{granick.s:2003,granick:02}. More
recent experimental data suggests that the description of flow near rough
surfaces has to be corrected, but for a separation, not
slip~\cite{vinogradova.oi:2006}. The theoretical description of such a flow
represents a difficult, nearly insurmountable, problem. It has been solved only
approximately, and only for a case of the periodic roughness and far-field flow
with a conclusion that it may be possible to approximate the actual surface by
a smooth one with the apparent slip boundary condition~\cite{sarkar.k:1996,lecoq.n:2004,kunert-harting-07}.

In this letter we address the fundamental, but still
open questions (i) whether the effect of random roughness on the flow may
be represented by replacing the no-slip condition on the exact boundary by
an effective condition on the equivalent smooth surface, (ii) where this
smooth surface is located, depending on geometric parameters of roughness,
and (iii) does this effective condition represent that of slip or no-slip? We will quite generally assume that the flow
near and far from the interface is a stable, laminar flow
field.
\begin{figure}[h]
\centerline{\hspace{0.0cm}\epsfig{file=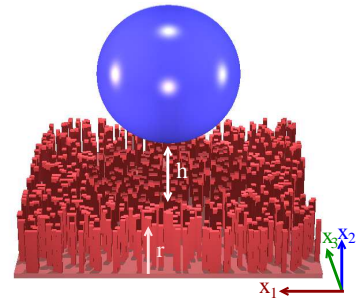,width=0.55\linewidth}}
\caption{ \label{fig:picture}
(Color online) Sketch of the system: a sphere of radius $R$ approaches a
rough surface with a fixed area fraction $\phi$ covered by roughness
elements. The separation $h$ is defined on top of the surface roughness at
position $x_2=r$.
}
\end{figure}

{\bf General idea and models.--} To address these issues we analyze
the hydrodynamic interaction between a smooth sphere of radius $R$ and a
rough plane (see Fig.~\ref{fig:picture}).  Beside its
significance as a geometry of SFA/AFM dynamic force experiments, this
allows us to explore both far and near-field flows in a single
``experiment''.
As an initial application we study roughness elements of a
fixed height $r$ that are distributed at random uncorrelated positions
with a given probability $\phi$. Such a surface mimics a situation
explored in recent
experiments~\cite{bonaccurso.e:2003,granick.s:2003,granick:02}. In
Cartesian coordinates ${\bf x}=(x_1,x_2,x_3)$, a separation $h$
is defined on top of the roughness, $x_2=r$, which finds its definition in
the AFM experiment~\cite{bonaccurso.e:2003,vinogradova.oi:2006}.

The exact solution, valid for an arbitrary separation, for a sphere
approaching a smooth plane is given by theoretical solutions of Brenner
and Maude~\cite{Brenner:1961,Maude:1961},
\begin{eqnarray}
\label{eq:maudea}
&\!\!\!\!\!\!\!\!\!\!\!\frac{F_1}{F_{St}}=-\frac{1}{3} \sinh \xi  \\
&\!\!\!\!\!\!\!\!\!\!\!\times\left(\sum^{\infty}_{n=1} \frac{n(n+1)\left [8e^{(2n+1)\xi}+2(2n+3)(2n-1)\right ]}{(2n-1)(2n+3)[4\sinh^2(n+\frac{1}{2})\xi-(2n+1)^2\sinh^2\xi]}\right.\nonumber\\
&\!\!\!\!\!\!\!\left.-\sum^{\infty}_{n=1}
\frac{n(n+1)\left [ (2n+1)(2n-1)e^{2\xi}-(2n+1)(2n+3)e^{-2\xi} \right ]}{(2n-1)(2n+3)[4
\sinh^2(n+\frac{1}{2})\xi-(2n+1)^2\sinh^2\xi]}\right),\nonumber
\end{eqnarray}
with $F_{St}=6 \pi \mu R v$, where $\mu$ is the dynamic viscosity, $v$ is
the velocity, and $\cosh \xi=h/R$, $\xi<0$. The leading term of this
expression can be evaluated as
\begin{equation}\label{firstorder}
    \frac{F_2}{F_{St}}\sim 1+\frac{9}{8}\frac{R}{h}.
\end{equation}
At large separations, $h \gg R$, the hydrodynamic force on a sphere turns
to the Stokes formula, but at small distances, $h \ll R$,  the drag force
is inversely proportional to the gap, $F_2/F_{St} \to 9 R/(8 h)$. A
consequence of this lubrication effect is that the sphere would never
touch the wall in a finite time.
The flow in the vicinity of a rough surface
should deviate from these predictions. A possible assumption is
that the boundary condition at the plane $x_2=r$ should be written as a
slip condition~\cite{bonaccurso.e:2003}.
To investigate this scenario we suggest to present a force as a product of
Eq.~\ref{firstorder} and a correction for slip
\begin{equation}\label{firstorder_slip}
    \frac{F_3}{F_{St}}\sim \left(1+\frac{9}{8}\frac{R}{h}\right) f^{\ast},
\end{equation}
where this correction, $f^{\ast}$, is taken to be equal as predicted for a lubrication force
between a no-slip surface and a surface with partial slip~\cite{vinogradova:95}.
\begin{equation}\label{model2}
   f^{\ast} = \frac{1}{4} \left( 1 + \frac{3 h}{2 b}\left[ \left( 1 +
\frac{h}{4 b} \right) \ln \left( 1 + \frac{4 b}{h} \right) - 1
\right]\right).\nonumber
\end{equation}
Another assumption would be that the rough surface is hydrodynamically
equivalent to a smooth one located somewhere between the top and bottom of
rugosities (at $x_2=r_{\rm eff}=r-s$). As found
in~\cite{vinogradova.oi:2006}), the force can be represented as
\begin{equation}
\label{approximation}
    \frac{F_4}{F_{St}}\sim 1+\frac{9}{8}\frac{R}{h+s}.
\end{equation}
At small $h$ expressions \ref{firstorder_slip} and
\ref{approximation} give different asymptotic behavior of a drag force,
$F_3/F_{St} \to 9R/(32 h)$, and  $F_4/F_{St} \to 9 R/(8 s)$. While
the second scenario allows a sphere to touch a plane, in the
first model this is impossible since the drag force diverges (but
differs from the standard lubrication asymptotics by a factor of 4).
Thus, a drainage study allows to distinguish
between these two models of hydrodynamic flow past rough surfaces.

{\bf Simulation method.--}
We apply the lattice Boltzmann (LB) method to
simulate the flow field between a smooth sphere approaching a rough
plane~\cite{bib:higuera-succi-benzi,2002RSPTA.360..437D,ladd01}.
The method allows precise measurements of the force acting on the sphere
and to explore the very large range of parameters. Besides that, in our
simulations we can consider a ``clean'' situation of a hydrodynamic force
and avoid effects of surface forces which significantly complicate the
analysis of SFA/AFM data. Since the method is well established, we only
shortly describe it here. By using a discretized and linearized version of
Boltzmann's equation
\begin{equation}
n_i({\bf x}+{\bf c}_i,t+1) - n_i({\bf x},t) = \sum_{j}\Lambda_{ij}(n^{\rm
eq}_j-n_j({\bf x},t)),
\label{eq:LBE}
\end{equation}
the LB approach allows to fully resolve the hydrodynamics~\cite{bib:succi-01}.
Positions $\bf{x}$ are discretized on a 3D lattice with $19$ discrete
velocities ${\bf c}_i$ pointing to neighboring sites. Each ${\bf c}_i$
relates to a single particle distribution function $n_i({\bf x},t)$
which is advected to neighboring sites at every time step. Then,
$n_i({\bf x},t)$ is
relaxed towards a local equilibrium $n_i^{\rm eq}(\rho,{\bf j})$
with a rate given by the matrix elements $\Lambda_{ij}$.
Mass $\rho$ and momentum ${\bf j}$ as given by moments of
$n_i({\bf x},t)$ are conserved.
We use the natural units of the system, i.e. the lattice constant
$\delta{\rm x}$ for the length and the time step $\delta{\rm t}$ for time.
Massive particles are described by a continuously moving boundary which is
discretized on the lattice. Momentum from the particle to the fluid is
transfered such that the fluid velocity at the boundary equals the
particle's surface velocity. Since the momentum transferred
from the fluid to the particle is known, the hydrodynamic force
can be recorded.
If not stated otherwise, the $256^3\delta{\rm x^3}$ system contains a
sphere with radius $R=16\delta{\rm x}$ which is moved in $y$ direction
at constant velocity $v=10^{-3}\delta{\rm x}/\delta{\rm t}$. The fluid
density is kept constant and the kinematic viscosity is $\mu/\rho=0.1$
resulting in a Reynolds number $Re=0.16$.
No-slip surfaces are described by mid-grid bounce back boundaries and
a slip boundary is implemented by a repulsive mean-field force
acting between fluid and surface~\cite{bib:jens-kunert-herrmann:2005,jari-jens-08}.
We carefully checked the influence of system size,
radius, and separation to insure that finite size and resolution
effects are negligible~\cite{kunert-harting-09}. Also, by testing
different resolutions we assured that a lateral width of roughness
elements of $1\delta{\rm x}$ is sufficient.
\begin{figure}[h]
\includegraphics[width= 0.16\textwidth,angle=270,clip]{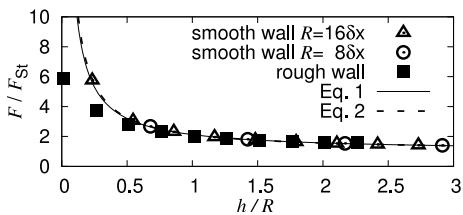}
\caption{\label{fig:forceflat}
Hydrodynamic force acting on a sphere with
radius $R=16\delta{\rm x}$ (triangles) and $R=8\delta{\rm x}$ (circles)
driven to a smooth wall with velocity $v=10^{-3}\delta{\rm x}/\delta{\rm t}$ and
$v=10^{-4}\delta{\rm x}/\delta{\rm t}$, correspondingly. The solid and
dashed curves are calculations of the force expected with no-slip boundary
conditions at the wall (Eqs.~\ref{eq:maudea}, \ref{firstorder}). Squares
are the results measured for a rough wall ($\phi=4\%, r=10\delta{\rm x}$).
}
\end{figure}

{\bf Results and discussion.--}
We test our method by measuring the hydrodynamic interaction between
smooth surfaces. Fig.~\ref{fig:forceflat} shows the normalized
hydrodynamic force for two simulation sets. In the first one, a sphere of
$R=8\delta{\rm x}$ is driven with $v=10^{-4}\delta{\rm x}/\delta{\rm t}$.
In the second run the sphere is twice as large, $R=16\delta{\rm x}$, and
the driving velocity is an order of magnitude larger, $v=10^{-3}\delta
{\rm x}/\delta{\rm t}$.
Fig.~\ref{fig:forceflat} includes the exact theoretical curve,
Eq.~\ref{eq:maudea}. The fit is excellent for all separations, indicating
that large shear rates do not induce any slip, a conclusion which does
not support recent experimental data~\cite{lauga2007}. Note that the
first-order approximation, Eq.~\ref{firstorder}, practically coincides
with the exact solution.
These simulations demonstrate that finite size effects and resolution
effects can be well controlled: for
$h<2R$ a $256^3\delta{\rm x^3}$ system is found to be sufficient to avoid
artefacts at large separations $h$~\cite{kunert-harting-09}.
Separations $<1\delta{\rm x}$ are excluded from the analysis since the
finite resolution leads to larger deviations.
Also included in Fig.~\ref{fig:forceflat} is a
normalized force measured near a rough wall ($\phi=4\%$, $r=10\delta{\rm
x}$), which at small distances is much smaller than predicted by
Eq.~\ref{eq:maudea}. This is qualitatively consistent with the AFM
observations~\cite{bonaccurso.e:2003,vinogradova:03}, but in contrast to
the SFA data~\cite{granick:02}, which likely reflects a different way of a
definition of zero separation in the SFA (at the bottom of asperities).
\begin{figure}
\includegraphics[width= 0.48\textwidth,angle=0,clip]{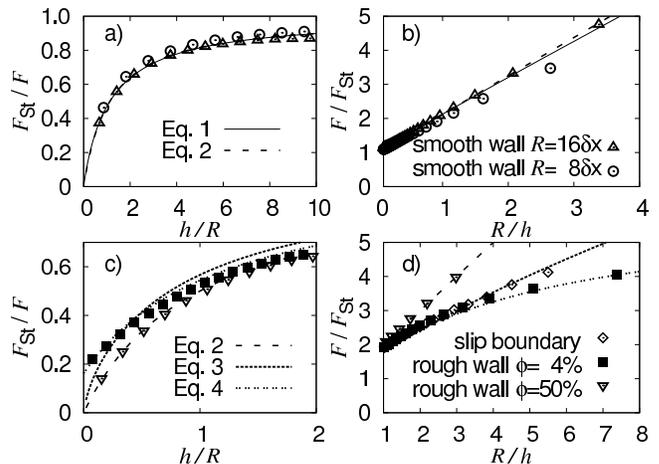}
\caption{ \label{fig:analysis1}
Hydrodynamic force plotted in different
coordinates. (a,b) the data sets for smooth surfaces reproduced from Fig.\ref{fig:forceflat},
(c,d) show the force for two rough planes with $r=10\delta{\rm x}$,
 $\phi=4\%$ and $\phi=50\%$. For $\phi=4\%$
the asymptotic behavior for small $h$ cannot be fitted with a slip, but the
assumption of an effective boundary position holds. The values for $\phi=50\%$
recover the case of a flat surface at $r$. The data for a smooth slip
boundary confirms the validity of
Eq.~\ref{firstorder_slip}.
}
\end{figure}

To examine the significance of roughness more closely, the force curves
from Fig.~\ref{fig:forceflat} are reproduced in Fig.~\ref{fig:analysis1}
in different coordinates. Figs.~\ref{fig:analysis1}a and b are intended to
indicate that both near field and far field theoretical asymptotics for
smooth surfaces are well reproduced in simulations.
Figs.~\ref{fig:analysis1}c and d show that simulation data for a rough
surface ($\phi=4\%$, $r=10\delta{\rm x}$) show deviations from the
behavior predicted
by Eq.~\ref{eq:maudea}. A possible explanation for this discrepancy is
that we invoke slippage at the wall, as modeled by
Eq.~\ref{firstorder_slip}.
This is illustrated in
Figs.~\ref{fig:analysis1}c and d, where the simulation data are compared
with another theoretical calculation in which a constant slip length of
$b=2.55\delta{\rm x}$, obtained from  the best possible fit of the force
curve, is incorporated in the model. This has the effect of decreasing the
force, and it provides a reasonable fit to the data down to $h/R \sim 3$,
but at smaller gap the simple model of slip fails to describe simulation
data, by predicting a larger force and its different asymptotic behavior.
This suggests, that it can only be considered as a first approximation,
valid at large distances from the wall. This conclusion is consistent with
early results obtained for a far field
situation~\cite{sarkar.k:1996,lecoq.n:2004,kunert-harting-07}, but does
not support recent AFM data~\cite{bonaccurso.e:2003}.
However, as shown by the simulation data, Eq.~\ref{firstorder_slip} is well applicable in the case of a slippery
wall.
An alternative
explanation for the smaller force compared to the theory for smooth
surfaces, Eq.~\ref{eq:maudea}, can be obtained if we assume that the
location of an equivalent effective wall, where no-slip boundary
conditions are applied, should be shifted, as modeled by
Eq.~\ref{approximation}. A corresponding theoretical calculation of the
drag force is shown in Figs.~\ref{fig:analysis1}c and d. This estimate
requires knowledge of the effective wall position $r_{\rm eff}$. The value
$r_{\rm eff}=7.86\delta{\rm x}$ was obtained from the fit of the measured
force curve and is enough to give a good match to the data at very small
distances, which confirms the conclusions of a recent
experiment~\cite{vinogradova.oi:2006}.
\begin{figure}[h]
\centerline{
\includegraphics[width=0.17\textwidth,angle=270,clip]{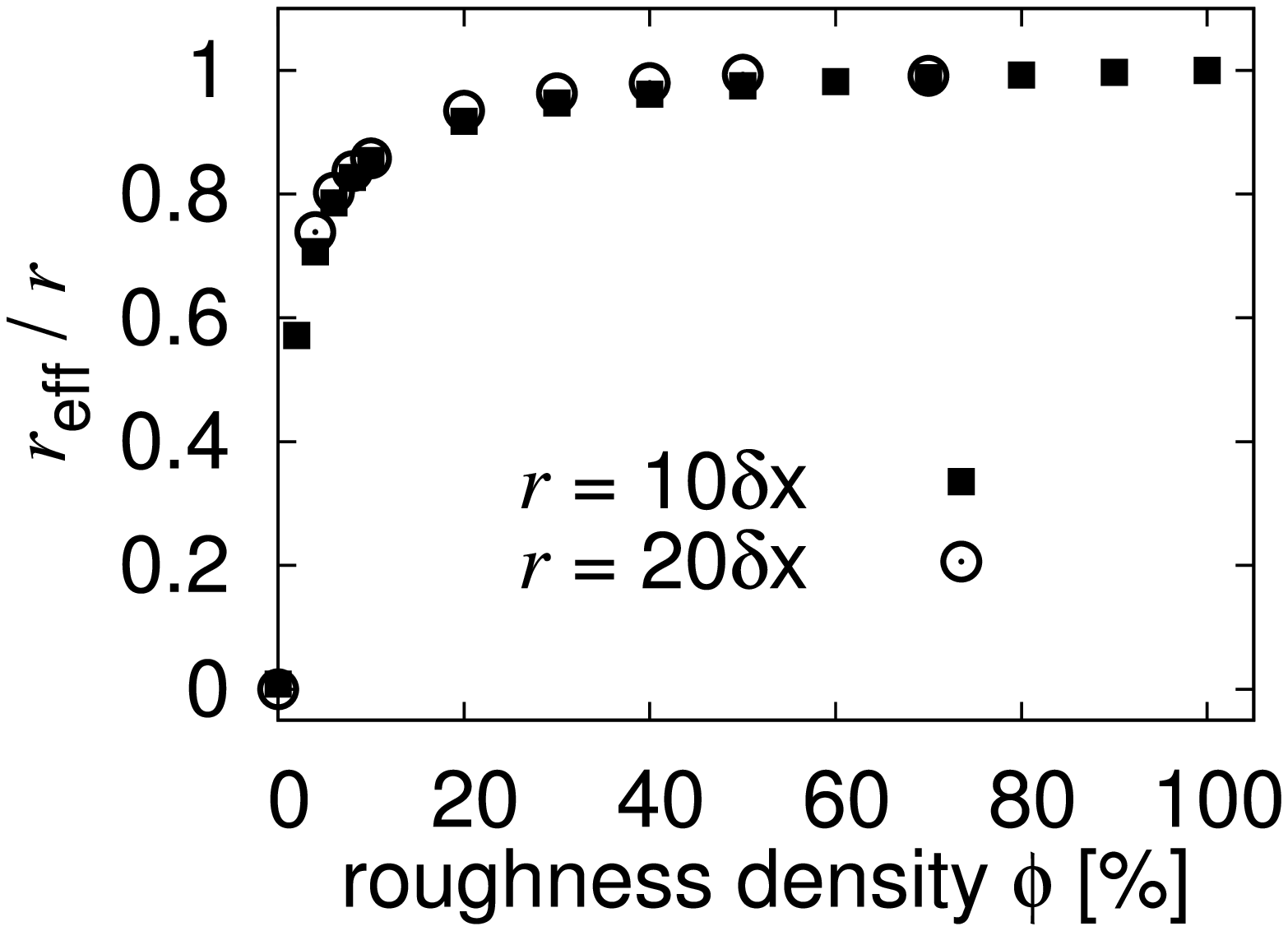}
\includegraphics[width=0.17\textwidth,angle=270,clip]{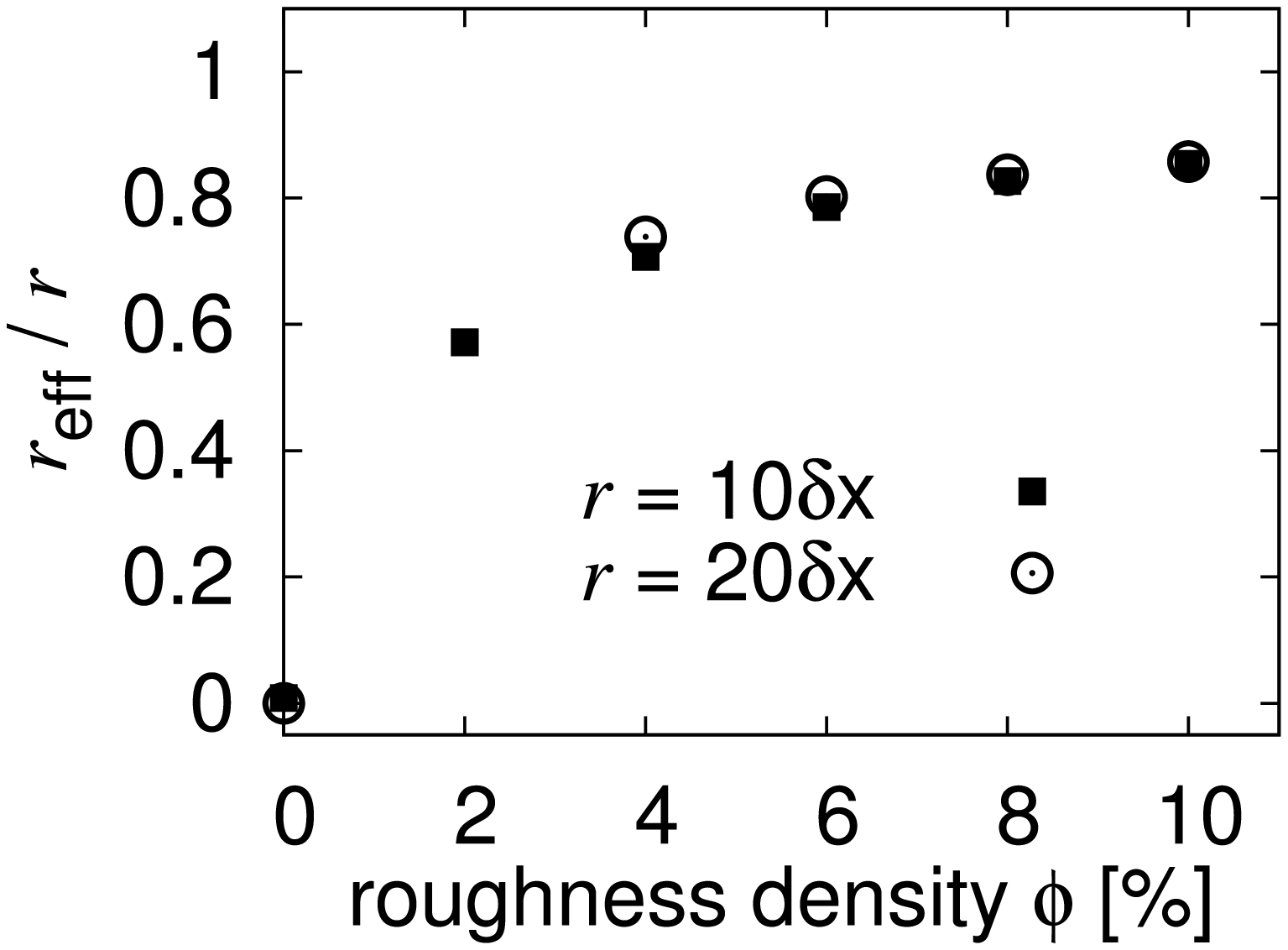}}
\caption{\label{fig:percent}
Effective height $r_{\rm eff}$ normalized by the maximum height $r$ as a
function of a density of roughness elements, $\phi$, for $r=10\delta{\rm
x}$ and $r=20\delta{\rm x}$ plotted in different scales.
}
\end{figure}

By performing similar fits for a variety of drainage runs with different $\phi$
and for surfaces with different height of roughness elements
($r=10\delta{\rm x}$ and $r=20\delta{\rm x}$) as well as its different
lateral width ($\delta{\rm x}$ and $2\delta{\rm x}$) we find that the
same conclusion is valid for all situations, but $r_{\rm eff}/r$ is itself
a function of $\phi$ (being surprisingly insensitive to the value of $r$).
In Fig.~\ref{fig:percent} we examine this in more detail. The simulation
data show that $r_{\rm eff}$ required to fit each run increases from 0 to
$r$ very rapidly, so that at $\phi=20\%$ it is already above $0.9 r$, and
at $\phi =50\%$ it is almost equal to $r$. This is illustrated by
including the data obtained for a larger density of roughness elements
($\phi=50\%$, $r=10\delta{\rm x}$) in Fig.~\ref{fig:analysis1}c, that do
not show a discernible deviation from the theoretical predictions for
smooth surfaces. Thus, a small number of roughness elements has enormous
influence on film drainage, confirming earlier theoretical
ideas~\cite{fan.th:2005}.

{\bf Conclusion.--}
We have presented lattice Boltzmann simulations describing the
drainage of a liquid confined between a smooth sphere and a randomly rough
plate. The measured force is smaller than predicted for two smooth
surfaces if the standard no-slip boundary conditions are used in the
calculation. What our results show, however, is that at small separations
the force is even weaker and shows different asymptotics than expected if
we invoke slippage at the smooth fluid-solid interfaces. To explain this
we use the model of a no-slip wall, located at an intermediate position
(controlled by the density of roughness elements) between top and bottom
of asperities. Calculations based on this model provide an excellent
description of the simulation data. Besides this, by proving a correctness
of this simple model to describe flow past a randomly rough surface, we
have suggested a validity of a number of simple formulas for a
hydrodynamic drag force. Although formally they can only be considered as
first-order approximations, their accuracy is confirmed by simulation. Our
results open the possibility of solving quantitatively many fundamental hydrodynamic
problems involving randomly-rough interfaces, including contact angle
dynamics, coagulation and more.

\begin{acknowledgments}
We acknowledge A.J.C.~Ladd for his hospitality (C.~Kunert) and access to his
simulation code, the DFG for financial support (grants Vi~243/1-3 and
Ha~4382/2-1), and SSC Karlsruhe for computing time.
\end{acknowledgments}


\begin{thebibliography}{10}

\bibitem{vinogradova:03}
O.~I. Vinogradova and G.~E. Yakubov.
\newblock {\em Langmuir}, 19:1227, 2003.

\bibitem{charlaix.e:2005}
C.~Cottin-Bizonne, B.~Cross, A.~Steinberger, and E.~Charlaix.
\newblock {\em Phys. Rev. Lett.}, 94:056102, 2005.

\bibitem{vinogradova.oi:2009}
O.~I. Vinogradova, K.~Koynov, A.~Best, and F.~Feuillebois.
\newblock {\em Phys. Rev. Lett.}, 102:118302, 2009.

\bibitem{vinogradova:99}
O.~I. Vinogradova.
\newblock {\em Int. J. Miner. Process.}, 56:31 -- 60, 1999.

\bibitem{barrat:99}
J.~L. Barrat and L.~Bocquet.
\newblock {\em Phys. Rev. Lett.}, 82:4671 -- 4674, 1999.

\bibitem{andrienko.d:2003}
D.~Andrienko, B.~D\"unweg, and O.~I. Vinogradova.
\newblock {\em J. Chem. Phys.}, 119:13106, 2003.

\bibitem{bib:jens-kunert-herrmann:2005}
J.~Harting, C.~Kunert, and H.~Herrmann.
\newblock {\em Europhys. Lett.}, 75:328, 2006.

\bibitem{lauga2007}
E.~Lauga, M.~P. Brenner, and H.~A. Stone.
\newblock In C.~Tropea, A.~Yarin, and J.~F. Foss, editors, {\em Handbook of
  Experimental Fluid Dynamics}, chapter~19, pp 1219--1240. Springer, NY, 2007.

\bibitem{bib:joseph_superhydrophobic:2006}
P.~Joseph, C.~Cottin-Bizonne, J.~M. Benoi, C.~Ybert, C.~Journet, P.~Tabeling,
  and L.~Bocquet.
\newblock {\em Phys. Rev. Lett.}, 97:156104, 2006.

\bibitem{ou2005}
J.~{Ou} and J.~P. {Rothstein}.
\newblock {\em Physics of Fluids}, 17:103606, October 2005.

\bibitem{feuillebois.f:2009}
F.~Feuillebois, M.~Z. Bazant, and O.~I. Vinogradova.
\newblock {\em Phys. Rev. Lett.}, 102:026001, 2009.

\bibitem{jari-jens-08}
J.~Hyv\"aluoma and J.~Harting.
\newblock {\em Phys. Rev. Lett.}, 100:246001, 2008.

\bibitem{sbragaglia-etal-06}
M.~Sbragaglia, R.~Benzi, L.~Biferale, S.~Succi, and F.~Toschi.
\newblock {\em Phys. Rev. Lett.}, 97:204503, 2006.

\bibitem{kusumaatmaja-etal-08b}
H.~Kusumaatmaja, M.~L. Blow, A.~Dupuis, and J.~M.Yeomans.
\newblock {\em Europhys. Lett.}, 91:36003, 2008.

\bibitem{bonaccurso.e:2003}
E.~Bonaccurso, H.-J. Butt, and V.~S.~J. Craig.
\newblock {\em Phys. Rev. Lett.}, 90:144501, 2003.

\bibitem{granick.s:2003}
S.~Granick, Y.~Zhu, and H.~Lee.
\newblock {\em Nat. Mater.}, 2:221 -- 227, 2003.

\bibitem{granick:02}
Y.~X. Zhu and S.~Granick.
\newblock {\em Phys. Rev. Lett.}, 88:106102, 2002.

\bibitem{vinogradova.oi:2006}
O.~I. Vinogradova and G.~E. Yakubov.
\newblock {\em Phys. Rev. E}, 73:045302(R), 2006.

\bibitem{sarkar.k:1996}
K.~Sarkar and A.~Prosperetti.
\newblock {\em J. Fluid Mech.}, 316:223, 1996.

\bibitem{lecoq.n:2004}
N.~Lecoq, R.~Anthore, B.~Cichocki, P.~Szymczak, and F.~Feuillebois.
\newblock {\em J. Fluid Mech.}, 513:247, 2004.

\bibitem{kunert-harting-07}
C.~Kunert and J.~Harting.
\newblock {\em Phys. Rev. Lett}, 99:176001, 2007.

\bibitem{Brenner:1961}
H.~Brenner.
\newblock {\em Chem. Eng. Sci.}, 16:242--251, 1961.

\bibitem{Maude:1961}
A.~D. Maude.
\newblock {\em British J. Appl. Phys.}, 12:293--295, 1961.

\bibitem{vinogradova:95}
O.~I. Vinogradova.
\newblock {\em Langmuir}, 11:2213 -- 2220, 1995.

\bibitem{ladd01}
A.~J.~C. Ladd and R.~Verberg.
\newblock {\em J. Stat. Phys.}, 104:1191, 2001.

\bibitem{2002RSPTA.360..437D}
D.~d'Humi{\`e}res, I.~Ginzburg, M.~Krafczyk, P.~Lallemand, and L.-S. Luo.
\newblock {\em Phil. Trans. R. Soc. Lond. A}, 360:437, 2002.

\bibitem{bib:higuera-succi-benzi}
F.~J. Higuera, S.~Succi, and R.~Benzi.
\newblock {\em Europhys. Lett.}, 9:345, 1989.

\bibitem{bib:succi-01}
S.~Succi.
\newblock {\em The lattice {B}oltzmann equation for fluid dynamics and beyond}.
\newblock Oxford University Press, 2001.

\bibitem{kunert-harting-09}
C.~Kunert and J.~Harting.
\newblock {\em Proceedings of MNF}, 2009.

\bibitem{fan.th:2005}
T.~H. Fan and O.~I. Vinogradova.
\newblock {\em Phys. Rev. E}, 72:066306, 2005.

\end{thebibliography}

\end{document}